\documentclass[namedreferences]{SolarPhysics} 
\usepackage[optionalrh,solaenum]{spr-sola-addons} 
\usepackage{graphicx}        
\usepackage{color}           
\usepackage{url}             

\newcommand{\beq}{\begin{equation}} 
\newcommand{\eeq}{\end{equation}} 
\newcommand{\bfig}{\begin{figure}} 
\newcommand{\efig}{\end{figure}} 
\newcommand{\ben}{\begin{enumerate}} 
\newcommand{\een}{\end{enumerate}} 
 
\newcommand{\mbf}[1]{ {\mathbf #1} }

\begin{document} 
\begin{article} 
\begin{opening} 
 
\title{Comment on ``Resolving the $180^\circ$ Ambiguity in Solar Vector 
  Magnetic Field Data: Evaluating the Effects of Noise, Spatial 
  Resolution, and Method Assumptions''} 
 
\author{Manolis K. Georgoulis$^{1,2}$} 
\runningauthor{Georgoulis} 
\runningtitle{Comment on the Ambiguity Resolution Paper of Leka et 
  al. (2009)} 
 
   \institute{$^{1}$ Research Center for Astronomy and Applied 
     Mathematics (RCAAM), Academy of Athens, 
     4 Soranou Efesiou Street, Athens, 
     Greece, GR-11527 email: \url{manolis.georgoulis@academyofathens.gr} \\
     $^{2}$ Marie Curie Fellow \\} 
\begin{abstract} 
In a recent paper, Leka {\it et al.} ({\it Solar Phys.} {\bf 260}, 83, 2009)
constructed a synthetic vector magnetogram representing a three-dimensional 
magnetic structure defined only within 
a fraction of an $\textrm{arcsec}$ in height. 
They rebinned the magnetogram to simulate 
conditions of limited spatial resolution and then   
compared the results of various azimuth disambiguation methods on the 
resampled data. Methods relying on the  
physical calculation of potential and/or 
non-potential magnetic fields failed in nearly the same, extended 
parts of the field of view and \inlinecite{leka_etal09} attributed these 
failures to the limited spatial resolution. This study shows that the failure
of these methods is {\it not} due to the limited spatial resolution but due to
the narrowly defined test data. Such narrow 
magnetic structures are not realistic in the real Sun. Physics-based
disambiguation methods, adapted for solar magnetic fields extending to
infinity, are not designed to handle such data; hence, they could only fail
this test. I demonstrate how an 
appropriate limited-resolution disambiguation test can be 
performed by constructing a synthetic vector magnetogram very similar to 
that of \inlinecite{leka_etal09} but representing a structure defined  
in the semi-infinite space above the solar photosphere. For this 
magnetogram I find that even a simple potential-field disambiguation method
manages to resolve the ambiguity very successfully, regardless of limited
spatial resolution. Therefore, despite the conclusions
of \inlinecite{leka_etal09}, a proper 
limited-spatial-resolution test of azimuth disambiguation 
methods is yet to be performed in order to identify the 
best ideas and algorithms.  
\end{abstract} 
\keywords{Active Regions, Magnetic Fields; Instrumental effects; 
  Magnetic Fields, Photosphere; Polarization, Optical} 
\end{opening} 
 
\section{Introduction} 
     \label{S-Introduction}   
Properly resolving the azimuthal $180^\circ$-ambiguity in the 
transverse (perpendicular to the line-of-sight) component of 
solar vector magnetograms inferred by the Zeeman effect  
is a prerequisite for further exploiting these valuable data. 
Calculation of electric currents, magnetic energy and helicity budgets,
flow velocities via physical models, and most techniques of coronal
magnetic field extrapolation rely on disambiguated vector magnetograms. 
The azimuth ambiguity was realized 
at the dawn of Zeeman-based vector magnetography  
\cite{harvey_69} and continues to be an open research topic to 
this day. Decent-quality vector magnetograms  
used to be rare, even in the recent past.  
This situation is reversed nowadays 
with vector magnetograms routinely provided by the ground-based Vector 
SpectroMagnetograph (VSM; \opencite{henney_etal09}) of the Synoptic Optical 
Long Term Investigations of the Sun (SOLIS) facility \cite{keller_etal03} 
and by the 
space-based SpectroPolarimeter (SP; \opencite{lites_etal01}) of the Solar 
Optical Telescope (SOT; \opencite{tsuneta_etal08}) 
onboard the {\it Hinode} spacecraft. Vast amounts of 
seeing-free full-disk vector magnetograms are also anticipated by 
the Helioseismic and Magnetic Imager (HMI; \opencite{scherrer_etal02})  
onboard the {\it Solar Dynamics Observatory} (SDO) mission. 
For single-height magnetograms 
acquired either at photospheric or at chromospheric heights  
(see, {\it e.g.}, \opencite{leka_metcalf_03})  
azimuth disambiguation is an ill-posed problem: the height derivatives 
$(\partial / \partial z)$ of any parameter (other than the normal 
field component $B_z$ in case one uses the divergence-free condition 
$\nabla \cdot \mbf{B} =0$) are unknown. To tackle this problem, an   
array of disambiguation techniques of various sophistication levels 
have been proposed. For a detailed description of most of these 
methods see \inlinecite{metcalf_etal06} and references therein; 
in addition, \inlinecite{li_etal07} and \inlinecite{crouch_etal09}
investigated disambiguation based on the divergence-free condition {\it per se} 
and concluded independently that information in multiple heights is required
for the method to work. More generally, the assumptions adopted by  
each disambiguation method are the ones ultimately responsible for 
the quality of the disambiguation results.  
Precisely and self-consistently disambiguating a vector 
magnetogram is a formidable problem, especially when the studied magnetograms 
represent complex magnetic structures with a multipolar, stressed, or sheared 
photospheric or low chromospheric boundary.  
 
In a recent paper, \inlinecite{leka_etal09} (hereafter LE2009)
aimed to continue the seminal work of \inlinecite{metcalf_etal06} 
who evaluated the different disambiguation methods by comparing their results
on given (different between the two studies), 
synthetic vector magnetograms. The two studies resulted from a 
series of Azimuth Disambiguation Workshops held by the respective 
Working Group. Target vector magnetograms were synthetic 
because in this case there is an {\it a priori} known true
solution or ``answer'', against which one may compare different disambiguation
solutions. ``Answers'' are unknown for real solar data. 
While \inlinecite{metcalf_etal06} focused on 
complex, flux-imbalanced, but noise-free and fully-resolved magnetic 
structures, LE2009 focused on 
the effects of photon noise and limited spatial resolution in the 
disambiguation process. I believe that the effect of photon noise and 
its impact was treated fairly appropriately by LE2009. The 
chosen limited-resolution (``flowers'') magnetograms, 
however, exhibited a feature  
that effectively disabled most disambiguation methods:  
{\it the magnetic field vector was defined only within a 
narrow layer of  $0.18''$  above the perceived ``photosphere''}, 
{\it i.e.}, the plane on which 
disambiguation was tested. For any disambiguation method attempting 
physical calculations ({\it i.e.}, potential and/or  
non-potential magnetic field) the underlying assumption is that 
{\it magnetic structures extend to the semi-infinite space above the 
photosphere}. This is driven by the lack of 
knowledge of the structures' outer edges but it appears obvious  
that these structures extend well  
above the photosphere due to the decrease of the plasma 
density and the subsequent passing from the forced, possibly discontinuous,  
photospheric, to the force-free, space-filling, coronal 
magnetic fields ({\it e.g.}, \opencite{longcope_welsch_00}).  
Here I show that the assumption of a field 
defined only on and slightly above the photosphere makes it 
practically impossible to reproduce the ``answer'' field of 
LE2009 by any physics-based disambiguation method. 
Therefore, physics-based methods were subjected to a test that was 
not possible to handle, not due to the complexity of the synthetic data 
but, rather, due to the design of these data.  
Had the synthetic magnetogram been designed to extend well above the 
photosphere, the reported results of LE2009  
would have been very useful and revealing. But with such an unrealistic 
magnetic structure one must determine what really needs 
reconsideration: the test data that overlook fundamental physics of solar
magnetic fields, or the methods and models that are more adapted to the real
Sun.  
  
LE2009 did not discuss the narrow validity problem in
detail. Instead, they attributed the failure of most disambiguation methods to
correctly reproduce certain parts of the ``answer'' field to lost information
due to unresolved structure on the disambiguation plane. Here I show in two
different ways that the limited spatial resolution is {\it not} the reason of
these methods' failure: first, by disambiguating the original, fully-resolved
and unbinned magnetogram of LE2009. This test was not
undertaken in that paper. I find that physics-based methods
fail at practically the same areas as in limited-resolution
magnetograms, so failure cannot be attributed to the limited spatial 
resolution. Second, by constructing a semi-infinite magnetic structure and
its corresponding horizontal field on the disambiguation plane using
the normal field component of LE2009 as the boundary condition. 
In this case I find that 
even a conventional, potential-field disambiguation reproduces
both the full- and the limited-resolution ``answer'' fields much better than
what LE2009 reported{\footnote{Redoing the analysis for a semi-infinite
    magnetic structures was proposed to K. D. Leka and colleagues prior to
    publishing the LE2009 paper. These authors declined,
    which led to this work that undertakes this task.}}. 

In Section \ref{S-theory} I provide the theoretical background of 
finite- vs. semi-infinite volume magnetic 
structures. Section \ref{S-data} describes the semi-infinite 
synthetic data 
on which I test some disambiguation methods, along with the 
metrics quantifying the performance of these methods. Section 
\ref{S-disambig} describes my disambiguation results while in Section 
\ref{S-discuss} I discuss crucial aspects of physics- and optimization-based 
disambiguation methods. Closing remarks are presented in Section \ref{S-rem}.
\section{Theoretical Background} 
\label{S-theory} 
Assume a current-free, potential magnetic field $\mbf{B_{\textrm{P}}}$  
($\nabla \times \mbf{B_{\textrm{P}}}=0$) in a given, finite and bounded, 
volume $V$. \inlinecite{aly_87} showed 
that $\mbf{B_{\textrm{P}}}$ has a unique solution in $V$, fully constrained by 
the nonzero normal field component $B_n$ on the boundary $\partial V$ of the 
volume. For a semi-infinite volume bounded only by one (bottom) boundary 
(plane, sphere) on which $B_n \ne 0$ the problem becomes identical to 
assuming that infinity corresponds to a flux, or magnetic, surface where  
$B_n=0$. Therefore, $B_n$ at the bottom boundary can fully 
constrain $\mbf{B_{\textrm{P}}}$ above it  
\cite{schmidt_64,chiu_hilton77,alissandrakis_81,sakurai_82,gary_89}. 
Without loss of 
generality I assume a planar lower boundary so that $B_n$ 
is replaced by the opposite of the 
vertical magnetic field component $B_z$ on this 
plane (assumed isolated and hence infinite, surrounded by areas of 
zero magnetic field). Different potential-field solutions $\mbf{B'_{\textrm{P}}}$ 
applying to this planar boundary and a {\it finite} volume above it 
differ from the unique $\mbf{B_{\textrm{P}}}$ by a gauge $\nabla \psi$, where $\psi$ 
is a smooth scalar ($\nabla ^2 \psi =0$) constrained only by the 
normal-field condition on the finite $\partial V$: 
\beq 
\mbf{B'_{\textrm{P}}} = \mbf{B_{\textrm{P}}} + \nabla \psi\;\;. 
\label{Bp_deps} 
\eeq 
Unless information to constrain $\psi$ is available, it is practically 
impossible to determine $\mbf{B'_{\textrm{P}}}$ in the finite volume given the 
infinity of possible choices for $\nabla \psi$. As  
\inlinecite{sakurai_89} puts 
it, one must have a physical reason for choosing a finite volume.  
This is the core of the problem with the finite-size magnetic 
structure of LE2009: other than computational convenience, there 
are no physical reasons dictating its selection.  
Physics-based methods, however, follow the semi-infinite volume 
approach backed up by the well-known fact that 
strong-field magnetic structures observed in the photosphere 
extend well above it. Hence, any physics-based 
method attempting to reproduce 
$\mbf{B'_{\textrm{P}}}$ (say, a potential-field disambiguation method) will,
at best, 
reproduce $\mbf{B_{\textrm{P}}}$ which can be very different. In this case it is of
little meaning to attempt disambiguation because neither of the two possible
disambiguation solutions for $\mbf{B'_{\textrm{P}}}$ on a given plane can be
reproduced, or even guessed, by any calculation expecting a semi-infinite
magnetic structure. 

Assume now a non-potential, current-carrying magnetic field $\mbf{B}$ 
in the semi-infinite volume above the lower planar boundary. The 
current-carrying component $\mbf{B_{\textrm{c}}}$ will be superimposed to
$\mbf{B_{\textrm{P}}}$ in this case, so that  
\beq 
\mbf{B} = \mbf{B_{\textrm{P}}} + \mbf{B_{\textrm{c}}}\;\;. 
\label{Bnp} 
\eeq 
Notice that since $\mbf{B_{\textrm{P}}}$ is fully constrained by $B_z$ on the 
boundary, $\mbf{B_{\textrm{c}}}$ has only horizontal components on the boundary, 
{\it i.e.} $B_{c_z}=0$ \cite{georgoulis_05,georgoulis_labonte07}. The 
current-carrying component will be responsible for the electric 
current density $\mbf{J}$ on and above the boundary via Ampere's law  
\beq 
\mbf{J} = {{c} \over {4 \pi}} \nabla \times \mbf{B_{\textrm{c}}}\;\;. 
\label{amp} 
\eeq 
Consider now the field  
$\mbf{B'} = \mbf{B'_{\textrm{P}}} + \mbf{B'_{\textrm{c}}} =
\mbf{B_{\textrm{P}}} + \nabla \psi + \mbf{B'_{\textrm{c}}}$   
applying 
to a finite volume. This will respectively give rise to an electric 
current density $\mbf{J'}$ that can be very different from  
$\mbf{J}$. Both $\mbf{J}$ and $\mbf{J'}$ on the boundary  
have no dependence on $B_z$ because $B_{c_z}= B'_{c_z}=0$ on it.  
A disambiguation method aiming to, say, reproduce both 
$\mbf{B'_{\textrm{P}}}$ and $J'_z$ on the boundary  
by semi-infinite volume calculations 
will have an untenable task: at best, it will  
reproduce $\mbf{B_{\textrm{P}}}$ and $J_z$, respectively.  
In LE2009 the various disambiguation methods were  
asked to reproduce the finite-volume, non-unique   
$\mbf{B'_{\textrm{P}}}$ and $J'_z$ with only the information needed 
to reproduce the semi-infinite, unique $\mbf{B_{\textrm{P}}}$ and $J_z$.    
\section{Two Versions of the Synthetic ``Flowers'' Case and Comparison Metrics}  
\label{S-data} 
\subsection{``Flowers'' Cases for Finite and a Semi-Infinite Volumes} 
\label{S-flowers} 
The so-called ``flowers'' case was designed by LE2009 to represent a 
current-free magnetic structure in full resolution (Figure 4 of that 
paper). Therefore, the discussion about potential fields and 
Equation (\ref{Bp_deps}) of Section \ref{S-theory} applies here. The      
reader is referred to Section 3.2 of LE2009 for a detailed description  
of the ``flowers'' construction.  
I only reiterate here that the structure is 
semi-analytical and produced by constructing a potential field  
on two planes: a lower plane, where disambiguation is attempted, and a 
higher plane at distance equal to six pixel sizes, or $0.18''$ per the 
designed pixel size of $0.03''$. The narrowly spaced planes apparently 
allow some control of the potential-field solution within the thin 
layer in between without knowledge of the field on its lateral boundaries.  
 
To translate the original flowers case of LE2009 into a 
potential-field structure valid in the semi-infinite volume above the 
lower boundary, I have obtained the ``answer'' vertical field 
component $B_z$ (K.D. Leka and colleagues have made this available 
online{\footnote{\url{http://www.cora.nwra.com/AMBIGUITY_WORKSHOP/2006_workshop/FLOWERS/}}}) 
and extrapolated from it. To achieve zero currents, hence a 
potential-field solution, at machine accuracy I choose  
the accurate, but characteristically  
slow{\footnote{This run took about 24 days and 17 hours 
in a 16-core computing cluster.}} 
Green's functions method of \inlinecite{schmidt_64}:  
in particular, I use the classical definition of the 
potential field, $\mbf{B_{\textrm{P}}} = - \nabla \chi$, where $\chi$ is a smooth 
scalar. Then, if $\mbf{r}$ is the vector position on the lower boundary
($z=0$) and $(\mbf{r},z)$ is the resulting vector position in the 
semi-infinite space $z \ge 0$, \inlinecite{schmidt_64} showed that
\beq 
\chi (\mbf{r},z) = {{1} \over {2 \pi}} \int \int  
{{B_z (\mbf{r'}) dx' dy'} \over {\sqrt{(\mbf{r} - \mbf{r'})^2 + z^2} 
}}\;\;, 
\label{chi} 
\eeq 
where $\mbf{r'}=(x', y')$ and $\mbf{r} \ne \mbf{r'}$ for $z=0$. 
Calculation of $\mbf{B_{\textrm{P}}}$ for $z \ge 0$ becomes then straightforward. 
 
\bfig[t!] 
\centerline{\includegraphics[width=.9\textwidth,clip=]{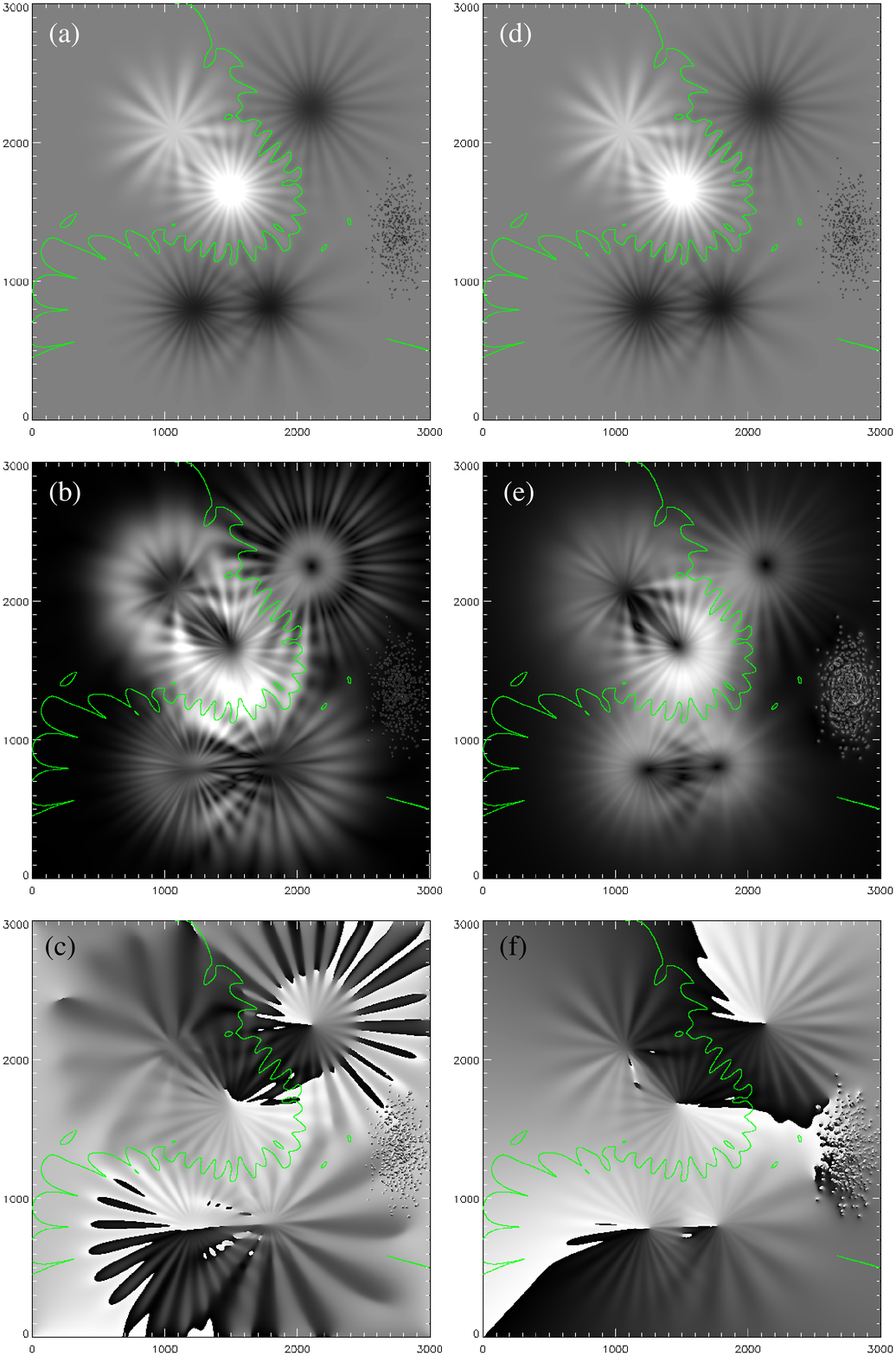}} 
\caption{Visual comparison between the field components of the 
  original, finite-volume flowers case of LE2009 (a-c) and my semi-infinite 
  volume flowers case (d-f) in full resolution. Shown are the identical 
  vertical field component saturated at $\pm 2.5$ kG (a,d), the horizontal 
  field strength saturated at $2.5$ kG (b,e), and the azimuth angle 
  (c, f), ranging between $0$ (black) and $2 \pi$ (white). The green 
  contour in all images indicates the location of the magnetic 
  polarity inversion line.} 
\label{viscomp} 
\efig 
At the lower (disambiguation) boundary, $z=0$, the original flowers case of
LE2009 and my semi-infinite flowers solution are depicted
in Figure \ref{viscomp}. For an identical vertical field component $B_z$ on the  
boundary (Figures \ref{viscomp}(a) and \ref{viscomp}(d)) 
there are very significant differences in both the horizontal field  
(Figures \ref{viscomp}(b) and \ref{viscomp}(e)) and the azimuth angle  
(Figures \ref{viscomp}(c) and \ref{viscomp}(f)). Both solutions give $J_z=0$ on 
the boundary, so they are potential-field solutions, but the flowers 
case of LE2009 cannot be reproduced unless the exact 
$B_z$-solution is known for the top boundary, as well (Equation (1) of  
LE2009). Obviously there 
are infinite potential-field solutions for the same $B_z$ on the lower 
boundary, each of them determined by the location and $B_z$-solution 
on the top and/or the lateral boundaries, but there is {\it only one} 
potential-field solution for the semi-infinite space.  
This solution is given in Figures \ref{viscomp}(d)-\ref{viscomp}(f). 
 
\bfig[t] 
\centerline{\includegraphics[width=.6\textwidth,clip=]{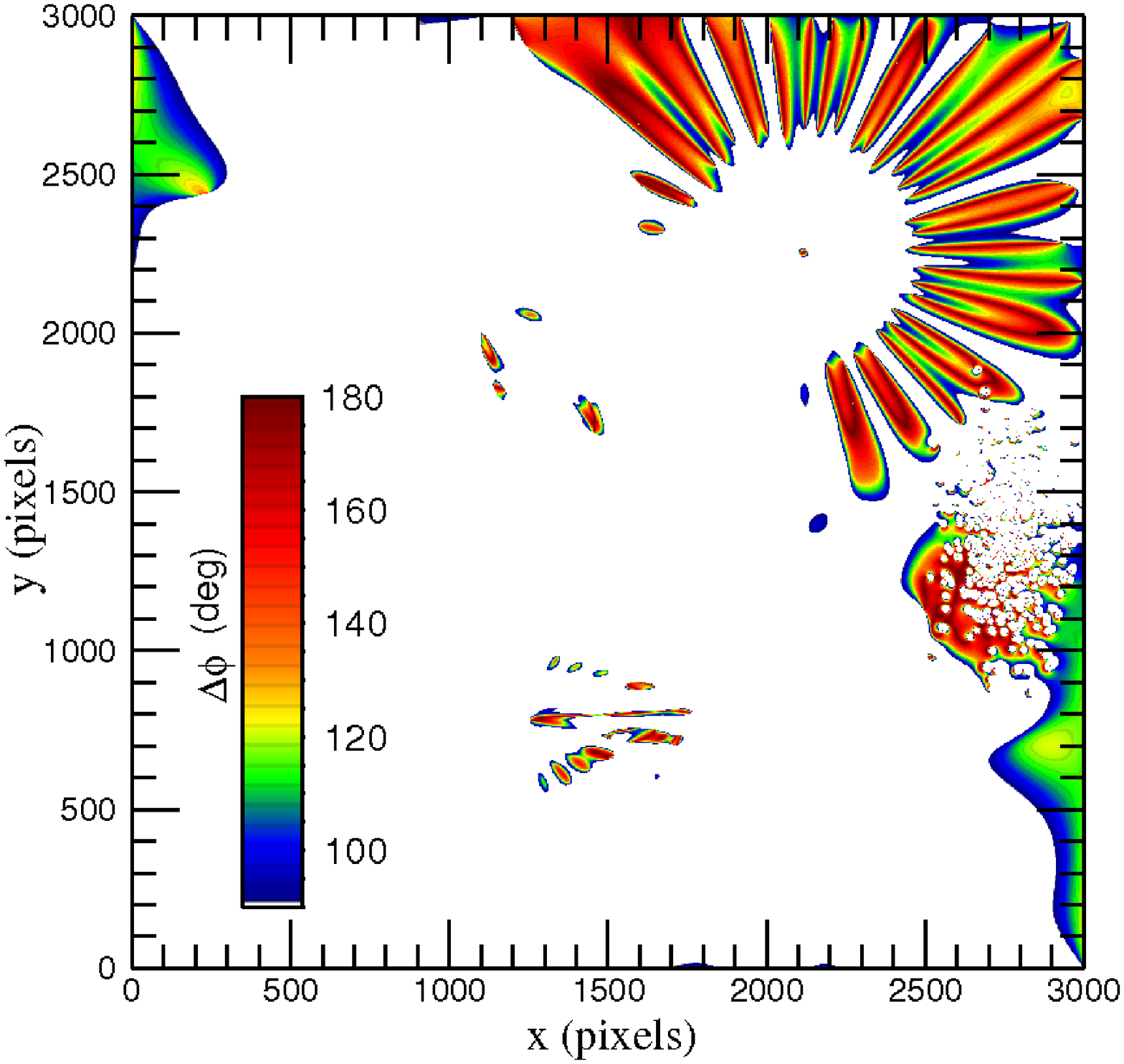}} 
\caption{Azimuth difference $\Delta \phi$ between the finite flowers 
  solution of LE2009 and my semi-infinite solution. Colored areas 
  correspond to $\Delta \phi > 90^o$.} 
\label{dphi} 
\efig 
It is revealing to compare the azimuth angles $\phi$ between the two flowers  
versions in full resolution (Figure \ref{dphi}). 
One notices that while the difference $\Delta \phi < 90^o$ 
for the majority of the boundary plane, there are extended areas where   
$\Delta \phi > 90^o$ (colored) or even $\Delta \phi \approx 180^\circ$  
(dark red). The latter correspond mostly to the ``ring of azimuth 
centers'' that LE2009 opted to simulate at the upper-right part 
of the structure and to the simulated ``plage'' area that extends 
between $x \in (2500, 3000)$ and $y \in (900, 1800)$. The colored 
areas in Figure \ref{dphi} are the ones for which at least an 
{\it acute-angle} potential-field disambiguation 
method is prone to fail. As I will show 
below, these are exactly the areas where much more sophisticated physics-based
methods invariably fail, as well. 
The reason of failure is not the difficulty to reproduce such 
complex fields or the limited resolution but merely the fact that 
these methods expect a semi-infinite magnetic structure, distinctly 
different from what they actually find.  
In other words, these areas are the ones 
most heavily impacted by the finite-volume approach. LE2009 argued that the 
simulated ring of azimuth centers and the plage can be descriptive of some 
solar-like structures. I am not countering this argument, 
but I question the overall concept of the flowers case as a 
finite-volume magnetic structure. 

Rebinning the full-resolution magnetogram to lower resolution, hence 
with a coarser pixel size, to simulate disambiguation in partially 
unresolved magnetic structures changes the ``answer'' field even 
more. LE2009 produced synthetic Stokes spectra and resulting 
images $I, Q, U, V$ and inverted them to obtain the ``answer'' field of 
Figures \ref{viscomp}(a)-\ref{viscomp}(c). Then they rebinned the spectra 
and inverted them 
to obtain the rebinned ``answer'' fields. This author is not able to 
fully reproduce this inversion process. Therefore, the 
semi-infinite flowers solution of Figures \ref{viscomp}(d)-\ref{viscomp}(f)  
has been spatially, rather than spectrally, rebinned.  
If anything, this rather simplistic spatial 
resampling might be expected to give more spurious structures when 
disambiguating the lower-resolution magnetograms.  
As we will see below, 
however, even simple rebinning of the data does not pose serious  
problems in the disambiguation.  
 
\bfig[t] 
\centerline{\includegraphics[width=1.\textwidth,clip=]{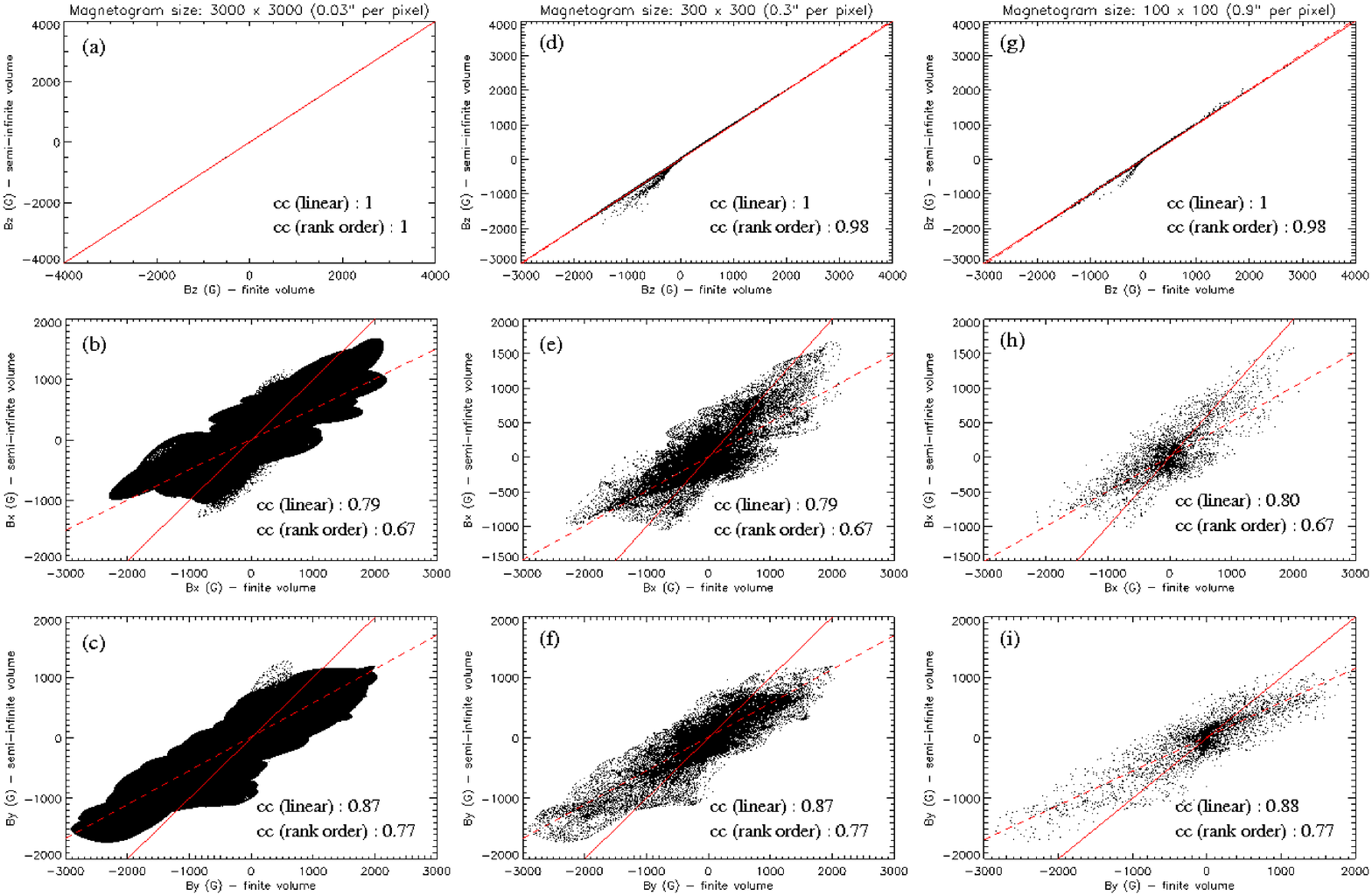}} 
\caption{Scatter-plot comparison between the magnetic field components 
  of my semi-infinite flowers solution (ordinate) and the original  
  flowers case of LE2009 (abscissa) at the lower boundary.  
  Shown are comparisons for 
  $B_z$ (a, d, g), $B_x$ (b, e, h), and $B_y$ (c, f, i) for the full 
  resolution (a-c), the $300 \times 300$-rebinned (d-f), and 
  the $100 \times 100$-rebinned (g-i) flowers cases. The solid 
  lines demonstrate equality and the 
  dashed lines show the least-squares best fit between any two 
  compared components. Also shown are the linear (Pearson)  
  and the non-parametric rank order (Spearman)  
  correlation coefficients for each comparison.} 
\label{scomp} 
\efig 
Scatter-plot comparisons between the magnetic field components 
for the original, finite-volume flowers cases of LE2009 
and the unique, semi-infinite volume flowers cases of this work  
are provided in Figure \ref{scomp}. Most of 
the differences correspond to $B_x$ and $B_y$, while the 
$B_z$-solutions in  
full-resolution (Figure \ref{scomp}(a)) are obviously identical. The 
small differences between the $B_z$-solution of LE2009 and my  
solution for the rebinned magnetograms (Figure \ref{scomp}(d) and 
\ref{scomp}(g)) refer primarily to the ``plage'' area -  
there, indeed, simple spatial rebinning incurs an impact on 
the lower-resolution $B_z$-maps.   
 
The scatter plots 
of Figure \ref{scomp} exhibit two notable features: {\it first}, 
correlation coefficients between any component $B_x$, $B_y$, $B_z$ for 
the original and the semi-infinite flowers solutions are nearly  
constant and insensitive to the rebinning process.  
This is strong indication that simple spatial rebinning does not 
introduce many more artifacts than the spectral rebinning and subsequent
inversion of LE2009, which is encouraging for this  
test. {\it Second}, the flowers cases of LE2009 
in both full and limited resolution invariably give stronger 
horizontal field  
components than the semi-infinite volume flowers case. These 
stronger fields cannot possibly be reproduced by a potential-field 
method seeking a minimum-energy field valid in the semi-infinite space.  
 
An important point here (that LE2009 also point out)  
is that the rebinned flowers magnetograms 
(either in the finite or in the semi-infinite volume) do not 
correspond to a potential-field solution any more. 
The loss of information due to rebinning spoils the smoothness of the 
scalar potential $\chi$ ($\nabla ^2 \chi$ now becomes nonzero - 
Section \ref{S-flowers}), 
thus incurring spurious currents that any disambiguation method has to
deal with. For the rebinned magnetograms 
the discussion about non-potential fields and Equations (\ref{Bnp}) and 
(\ref{amp}) of Section \ref{S-theory} apply. Overall, one should look  
critically on observed solar vector magnetograms as part of the inferred
electric currents may be due to the incomplete 
recording of the fine structure of the actual fields. The reader 
is referred to \inlinecite{parker_96} for an 
argument that all vertical current $J_z$ inferred from photospheric 
vector magnetograms is, in fact, fictional and caused by the 
limited resolution of the observing instruments. Further discussion  
exceeds the scope of this study. 
\subsection{Comparison Metrics} 
\label{S-metrics} 
As in LE2009, the quality of the performed 
disambiguations will be judged 
by an array of different metrics, each highlighting a certain aspect 
of the disambiguation solution. In particular, I use the following metrics:  
\ben 
\item[(1)] The {\it area metric}, $M_S$ (denoted by $M(a,s)_{\textrm{area}}$ 
  in LE2009): take the ratio of the number of pixels 
  $N_{\textrm{pixels}_{(\textrm{correct})}}$, where ambiguity has been resolved correctly, 
  over the total number of pixels $N_{\textrm{pixels}_{(\textrm{tot})}}$: 
\beq 
M_S = { {N_{\textrm{pixels}_{(\textrm{correct})}}} \over {N_{\textrm{pixels}_{(\textrm{tot})}}} }\;\;. 
\label{Ms} 
\eeq 
Achieving a $M_S$-value closer to 1 implies better disambiguation 
results with perfect disambiguation reflected in $M_S = 1$.  
\item[(2)] The {\it transverse field metric}, $M_{B_t > \mathcal{T}}$ 
  (denoted by $M(a, s)_{B_{\perp} > \mathcal{T}}$ in LE2009): 
  define a threshold $\mathcal{T}$ in the transverse field strength 
  and calculate the ratio between the sum of transverse field 
  $B_{t_{(\textrm{correct})} > \mathcal{T}}$ above this 
  threshold where ambiguity has been resolved correctly over the 
  total transverse field $B_{t > \mathcal{T}}$  
  above the threshold $\mathcal{T}$: 
\beq 
M_{B_t > \mathcal{T}} = { {\sum (B_{t_{(\textrm{correct})} > \mathcal{T}})} \over  
                        {\sum (B_{t > \mathcal{T}}) } }  
\label{MBt} 
\eeq 
Here, again, a value closer to 1 implies better disambiguation results 
with $M_{B_t > \mathcal{T}} = 1$ implying perfect disambiguation.  
\item[(3)] The {\it mean vector field difference metric},  
$M_{\Delta \mbf{B}}(a, s)$: take the magnitude of the difference 
  between the disambiguation solution $\mbf{B}_{(s)}$ and the 
  ``answer'' field $\mbf{B}_{(a)}$, sum it over the disambiguation 
  plane, and normalize by the total number of pixels 
  $N_{\textrm{pixels}_{(\textrm{tot})}}$: 
\beq 
M_{\Delta \mbf{B}}(a, s) = { {\sum (|\mbf{B}_{(s)} - \mbf{B}_{(a)} |) } 
                            \over {N_{\textrm{pixels}_{(\textrm{tot})}}} }\;\;. 
\label{MDB} 
\eeq
This is a dimensional metric, providing the mean difference 
between the disambiguation solution and the ``answer'' field in 
magnetic field units. The smaller the value of  
$M_{\Delta \mbf{B}}(a, s)$ the better the disambiguation with  
$M_{\Delta \mbf{B}}(a, s) = 0$ implying perfect disambiguation.
\item[(4)] The {\it normalized electric current density metric}, 
  $M_{J_z}(a, s)$: take the vertical current density $J_{z_{(s)}}$ 
  inferred by the disambiguation solution and the ``answer'' vertical 
  current density $J_{z_{(a)}}$ to form the metric  
\beq 
M_{J_z}(a, s) = 1 - { {\sum (|J_{z_{(a)}} - J_{z_{(s)}} |)} \over  
               {2 \sum (|J_{z_{(a)}}|) } }\;\;, 
\label{Mjz} 
\eeq 
where $\sum (\;)$ corresponds to summation over all pixels on the 
disambiguation plane. Here also, better results are reached if  
$M_{J_z}(a, s)$ tends to 1 with perfect disambiguation reflected in  
$M_{J_z}(a, s) =1$.  
\item[(5)] The {\it total vertical current metric}, $M_I$: this is another 
  dimensional metric and corresponds to the absolute sum of the 
  disambiguation-inferred  $J_{z_{(s)}}$ over the disambiguation 
  plane:  
\beq 
M_I = \sum (| J_{z_{(s)}} |)\;\;. 
\label{MI} 
\eeq  
This metric is compared to the respective sum obtained for the 
``answer'' vertical current density $J_{z_{(a)}}$. 
\een 
\section{Disambiguation Attempts}  
\label{S-disambig}       
To disambiguate both the finite- and the semi-infinite versions 
of the flowers case I basically use the non-potential magnetic field 
calculation (NPFC) method of \inlinecite{georgoulis_05} as revised in 
\inlinecite{metcalf_etal06}. This is denoted as NPFC2 in the latter  
paper and in LE2009 but I call it  
NPFC hereafter. Briefly, the method uses Equation (\ref{Bnp}) for the 
potential and the non-potential fields, $\mbf{B_{\textrm{P}}}$ and $\mbf{B_{\textrm{c}}}$, 
respectively. It is iterative, producing interim $\mbf{B_{\textrm{P}}}$- 
and $\mbf{B_{\textrm{c}}}$-solutions, the former constrained by the interim 
$B_z$-solutions and the latter by the interim $J_z$-solutions. 
At the end of each iteration, each 
interim solution is replaced by the closest combination of the two 
possible disambiguation solutions for the heliographic field 
components on the image (observation) plane. 
The whole scheme converges to a stable, self-consistent 
disambiguation solution that is then translated to the line-of-sight 
reference system to determine the preferred orientation of the 
transverse field for each pixel. For another detailed description of 
the NPFC method and its pipeline  
application to SOLIS/VSM data, see \inlinecite{georgoulis_etal08}.  
 
The NPFC method makes only {\it one} 
assumption: that the vertical component $B_{c_z}$ of the 
current-carrying field $\mbf{B_{\textrm{c}}}$ is constant with height  
$(\partial B_{c_z} / \partial z = 0)$ on the disambiguation 
plane.  This is a reasonable assumption  
given that $B_{c_z} =0$ on the disambiguation plane (see discussion in 
Section \ref{S-theory}) and that one generally expects $B_{c_z}$ to be 
small immediately above the plane unless the orientation of the 
magnetic field lines changes drastically. 
 
Since the NPFC method reconstructs the magnetic field on the boundary 
by the superposition of $\mbf{B_{\textrm{P}}}$ and $\mbf{B_{\textrm{c}}}$, one can always 
enforce $\mbf{B_{\textrm{c}}} =0$ on the boundary. This degenerates the NPFC 
method into a simple potential-field disambiguation method with 
$\mbf{B_{\textrm{P}}}$ calculated using fast Fourier transforms (FFT) for 
computational convenience.  
 
Here I will use both the potential (FFT) and the NPFC disambiguation 
methods. Potential-field methods constitute the simplest 
possible disambiguation approach and are admittedly unrealistic for  
complicated magnetic structures. As such, their results are considered minimum 
standards. I use a potential-field method here (albeit somewhat more
sophisticated than an acute-angle potential-field method) to show 
that even this works in some limited-resolution conditions if the test data  
are appropriately constructed, that is, they correspond to the semi-infinite
magnetic field solution.
 
\bfig[t!] 
\centerline{\includegraphics[width=.93\textwidth,clip=]{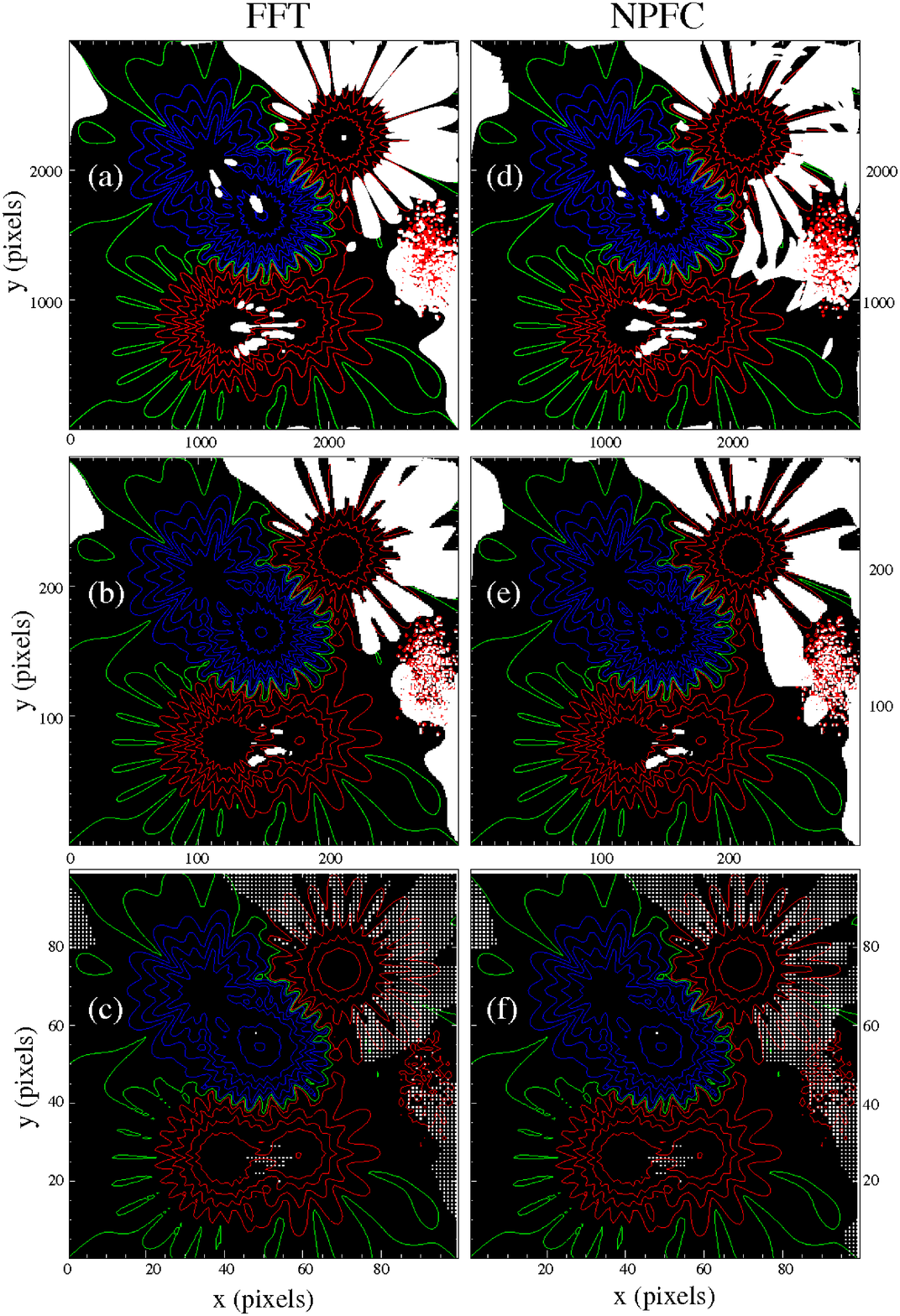}} 
\caption{Comparison of disambiguation solutions with 
  the ``answer'' field for the original flowers cases of LE2009.  
  Shown are potential-field (FFT) disambiguations (a-c) 
  and NPFC disambiguations (d-f). Comparisons refer to 
  full resolution (a, d), the $300 \times 300$-rebinned 
  field (b, e), and the $100 \times 100$-rebinned field (c, 
  f). White (black) areas indicate where the ambiguity was 
  incorrectly (correctly) resolved.  
  The contours correspond to the line-of-sight 
  field components and are taken at 0 and    
  $\pm (100, 200, 600, 1000, 2000, 3000)$ G. Contours are blue
  (positive polarity), red (negative polarity), and green (magnetic polarity
  inversion line).}
\label{dis_leka} 
\efig 
Figure \ref{dis_leka} depicts the FFT and the NPFC 
disambiguation solutions on the original, finite-volume flowers case 
of LE2009. Contrary to that paper, here I also disambiguate 
the fully resolved flowers case ($0.03''$ per pixel - 
Figures \ref{dis_leka}(a) and \ref{dis_leka}(d)) besides the two partially 
resolved cases: $0.3''$ per pixel (Figures \ref{dis_leka}(b) and  
\ref{dis_leka}(e)) and $0.9''$ per pixel  
(Figures \ref{dis_leka}(c) and \ref{dis_leka}(f)). The limited-resolution cases 
were disambiguated from the ambiguous magnetograms that LE2009  
originally made available for the tests. For the fully resolved 
case I randomly scrambled the azimuth of the ``answer's'' transverse 
field to create a $180^\circ$-ambiguous magnetogram.  
 
Both the potential (Figures \ref{dis_leka}(b) and \ref{dis_leka}(c)) 
and the NPFC (Figures \ref{dis_leka}(e) and (f)) disambiguation solutions 
generally reproduce the results of LE2009 
(the NPFC solution is present in LE2009 because one of the
authors used the NPFC algorithm available  
online{\footnote{\url{http://astro.academyofathens.gr/people/georgoulis/codes/ambiguity_resolution}}}).  
Considering also the disambiguation of the 
fully resolved data reveals a key finding in 
Figure \ref{dis_leka}: regardless of spatial resolution (full or limited), 
both the 
potential and the NPFC methods fail (white areas) at nearly the same 
parts of the field of view, 
{\it namely the areas where  $\Delta \phi > 90^o$ in Figure \ref{dphi}}, 
which are the ones most 
heavily impacted by the finite-volume approach.   
Therefore, failure in Figures  
\ref{dis_leka}(b), \ref{dis_leka}(e), \ref{dis_leka}(c), and \ref{dis_leka}(f) 
cannot be due to the limited spatial resolution, 
contrary to what LE2009 concluded.  
 
\bfig[t!] 
\centerline{\includegraphics[width=.93\textwidth,clip=]{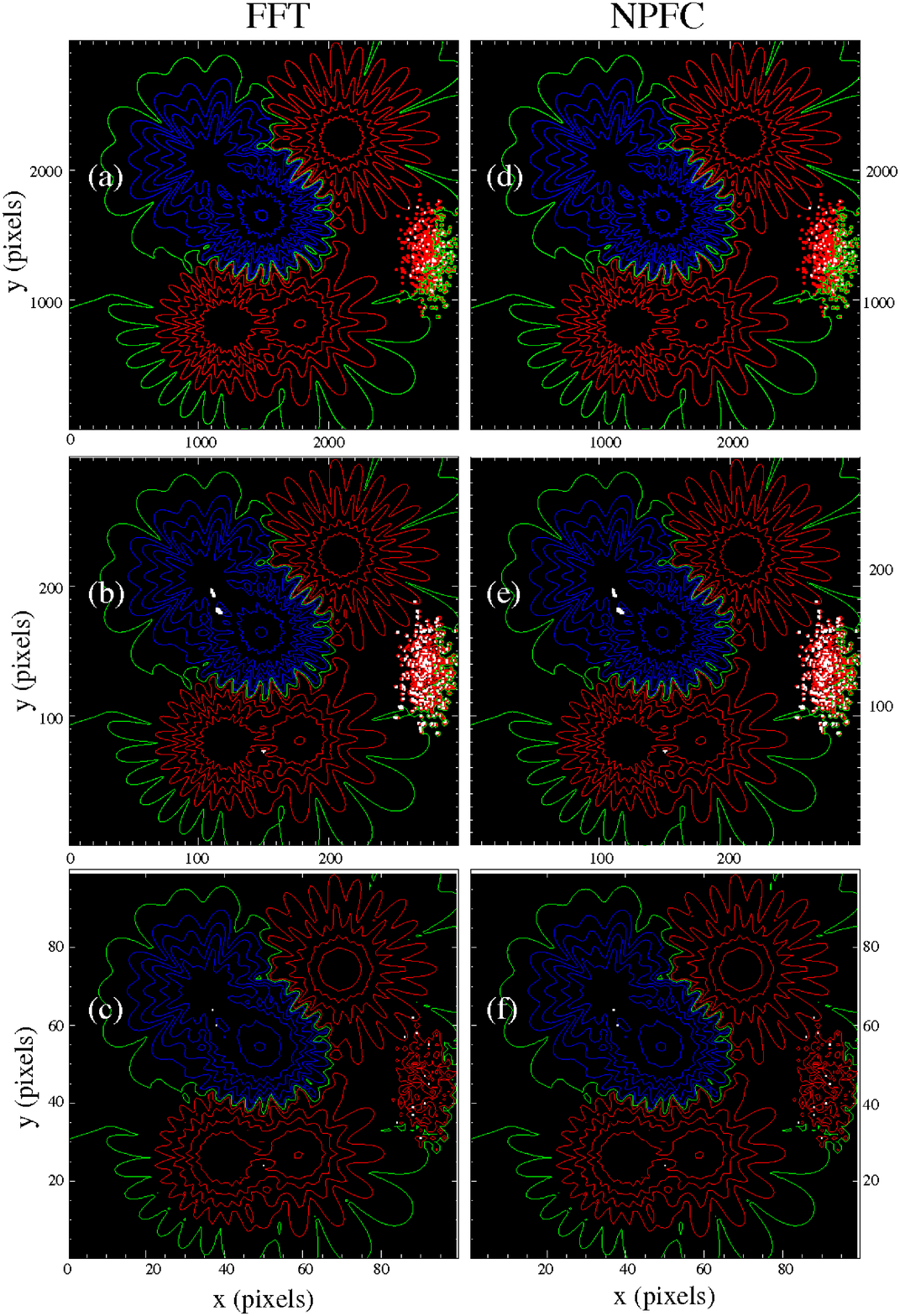}} 
\caption{Same as Figure \ref{dis_leka}, but for the semi-infinite 
  volume flowers case introduced in this work.}  
\label{dis_mkg} 
\efig 
Disambiguation tests on the semi-infinite-volume flowers cases are 
attempted in Figure \ref{dis_mkg}. Here I scrambled the azimuth in both 
the full- and the limited-resolution ``answers'' to construct the 
ambiguous data. One now sees that the ambiguity is resolved correctly 
in the vast majority of pixels {\it regardless} of full or limited  
spatial resolution and for {\it both} the potential and the NPFC 
methods. Minor inconsistencies (white areas) refer exclusively to the 
``plage'' area in the case of full resolution. For limited resolution,  
problems in the ``plage'' area seem to be enhanced, at least for the 
$0.3''$-per-pixel case (Figures \ref{dis_mkg}(b) and \ref{dis_mkg}(e)) 
and, in addition, 
some minor problems occur in areas of strong gradients in the magnitude and 
orientation of the transverse field (Figure \ref{viscomp}(e))  
due to the lost structure. Problems in the plage area and elsewhere 
may be either due to the simple rebinning or, indeed, due to spurious 
effects caused by the loss of spatial resolution. This being said,  
Figure \ref{dis_mkg} undoubtedly shows that  
{\it both} the potential and the NPFC methods correctly 
resolve the ambiguity in most of the field of view, {\it regardless} of 
full or limited spatial resolution, when the magnetic structure 
corresponds to the semi-infinite volume above the boundary.  
 
From Figures \ref{dis_leka} and \ref{dis_mkg}  
one notices that the finite-volume construction of the 
original flowers case, not the limited spatial resolution, is  
responsible for the problems in disambiguating these magnetograms.  
Methods that attempt to reproduce solar magnetic fields will fail by design  
to reproduce a field defined only within a narrow 
layer. On the other hand, such a structure is by no means expected in 
the solar atmosphere, where sunspot magnetic fields 
obviously extend well above the photosphere. Therefore, the original 
flowers case is not a proper test case for solar magnetic field  
disambiguation. Had LE2009 adopted a more ``solar-like'' semi-infinite 
structure they would have reached different results without having to 
sacrifice in fine detail: they could have produced their 
$B_z$ solution using the two-planes approach and then calculate the 
horizontal field pertaining to the semi-infinite volume above the 
lower boundary. LE2009 do make efforts to 
justify that their flowers magnetogram is ``solar-like'' (see their 
Figure 9 and subsequent discussion). I have attempted similar tests with my 
semi-infinite flowers case obtaining similar results. In this sense, the
semi-infinite flowers case is also a ``solar-like'' structure. The fact,
however, that the field of LE2009 is defined only within $0.18''$ above the
photosphere could be much less justified in a solar context.  
 
\begin{table}[t] 
\caption{Results of the first three comparison metrics ($M_S$, 
  $M_{J_z}(a, s)$, $M_I$) for the original (finite-volume) 
  flowers cases of LE2009 and for the semi-infinite-volume flowers 
  cases introduced here. The metric values are rounded in their 
  closest second decimal digit. I test a potential-field (FFT) and 
  the NPFC disambiguation method. Notice that the $M_{J_z}(a, s)$-values in full
  resolution in the semi-infinite flowers case are not given because
  $J_{z_{(a)}}=0$ almost up to machine accuracy in this case (see also $M_I$
  in full resolution), so the metric cannot be defined (Equation (\ref{Mjz})).} 
\label{tb1} 
\begin{tabular}{lccccccccccc} 
\hline 
               & \multicolumn{11}{c}{Original (finite-volume) flowers cases of LE2009} \\ 
               & \multicolumn{3}{c}{$M_S$} &   
               & \multicolumn{3}{c}{$M_{J_z}(a, s)$} &   
               & \multicolumn{3}{c}{$M_I$ ($\times 10^{13}\;A$)} \\ 
\cline{2-4}\cline{6-8}\cline{10-12}\\ 
               & $0.03''$ & $0.3''$ & $0.9''$ & & $0.03''$ & $0.3''$ & $0.9$'' & & $0.03''$ & $0.3''$ & $0.9''$ \\ 
\hline 
``Answer''     &      &      &      & &        &      &      & & 0.02 & 0.49 & 0.90\\ 
Potential (FFT)& 0.84 & 0.85 & 0.85 & & -18.57 & 0.41 & 0.74 & & 0.84 & 0.90 & 0.98\\ 
NPFC           & 0.84 & 0.85 & 0.86 & & -29.44 & 0.42 & 0.41 & & 1.30 & 0.92 & 1.16\\ 
\hline 
               & \multicolumn{11}{c}{Semi-infinite-volume flowers cases introduced here} \\ 
\hline 
``Answer''     &      &      &      & &      &      &     & & 0.00 & 0.39 & 0.39\\ 
Potential (FFT)& 1.00 & 1.00 & 1.00 & & N/A & 0.88 & 1.00 & & 0.00 & 0.44 & 0.39\\ 
NPFC           & 1.00 & 1.00 & 1.00 & & N/A & 0.87 & 1.00 & & 0.00 & 0.44 & 0.39\\ 
\hline 
\end{tabular} 
\end{table}  
\begin{table}[h] 
\caption{Same as Table \ref{tb1} but for the remaining metrics  
         $M_{B_t > \mathcal{T}}$ and $M_{\Delta \mbf{B}}(a, s)$.}  
\label{tb2} 
\begin{tabular}{lccccccccccc} 
\hline 
               & \multicolumn{11}{c}{Original (finite-volume) flowers cases of LE2009} \\ 
               & \multicolumn{3}{c}{$M_{B_t > 100\;\textrm{G}}$ } &  
               & \multicolumn{3}{c}{$M_{B_t > 500\;\textrm{G}}$} &  
               & \multicolumn{3}{c}{$M_{\Delta \mbf{B}}(a, s)$ (G)}\\ 
\cline{2-4}\cline{6-8}\cline{10-12}\\ 
               & $0.03''$ & $0.3''$ & $0.9''$ & & $0.03''$ & $0.3''$ & $0.9''$ & & $0.03''$ & $0.3''$ & $0.9''$ \\ 
\hline 
Potential (FFT)& 0.94 & 0.94 & 0.94 & & 0.98 & 0.99 & 0.98 & & 72.66 & 65.87 & 61.70\\ 
NPFC           & 0.93 & 0.94 & 0.95 & & 0.97 & 0.98 & 0.98 & & 78.44 & 66.18 & 64.88\\ 
\hline 
               & \multicolumn{11}{c}{Semi-infinite-volume flowers cases introduced here} \\ 
\hline 
Potential (FFT)& 1.00 & 1.00 & 1.00 & & 1.00 & 1.00 & 1.00 & & 0.00 & 2.24 & 0.20\\ 
NPFC           & 1.00 & 1.00 & 1.00 & & 1.00 & 1.00 & 1.00 & & 0.00 & 2.26 & 0.20\\ 
\hline 
\end{tabular} 
\end{table} 
Tables \ref{tb1} and \ref{tb2} provide the results of the 
comparison metrics introduced in Section \ref{S-metrics}. While these 
results for the original flowers case of LE2009 are largely 
consistent with those included in that paper for the NPFC and the 
potential-field methods, the results for the semi-infinite flowers case  
introduced here demonstrate that the disambiguation has been tackled much 
more efficiently by both methods {\it regardless} of limited 
spatial resolution and even the simple rebinning.   

A passing note should address the plausible question of why the simplistic
potential-field disambiguation scores so highly in the semi-infinite flowers
case, identically to (and, in a couple of cases, slightly better than) 
the more sophisticated NPFC method. The answer
lies in the properties of the test case. Despite its structural complexity,
flowers is a potential-field model that 
does not involve strong polarity inversion lines, significant
magnetic stresses, or strong shear. If it did, as shown clearly in
\inlinecite{metcalf_etal06}, (acute-angle or not) potential-field
disambiguation would largely fail in these areas. 
\section{Discussion} 
\label{S-discuss} 
\subsection{Optimization vs. Physics-Based Methods} 
\label{S-debate} 
Two of the tested disambiguation methods in LE2009 appeared to be 
less sensitive to the limited-resolution flowers cases than methods 
calculating potential / non-potential fields: the AZAM utility, 
implemented by B. W. Lites, and the ``minimum-energy'' (ME0) method, 
which is a revisit of \citeauthor{metcalf_94}'s \shortcite{metcalf_94} 
``minimum energy'' method. 
The AZAM 
is a non-automatic disambiguation method and requires 
a human operator. It 
enforces smoothness on the disambiguation results along with an 
empirical compliance with the divergence-free condition. The results, 
therefore, are largely subject to the operator's skill. The AZAM 
method has managed to reproduce some of the best disambiguation 
solutions in tests but since it is not automatic it cannot be part 
of massive disambiguation efforts. The 
ME0 method is an optimization and uses simulated annealing to minimize 
the functional  
\beq 
\mathcal{F} = |\nabla \cdot \mbf{B}| + \lambda |J_z| 
\label{me0} 
\eeq 
over the disambiguation plane. The choice for the weighting factor 
$\lambda$ changes  
the significance of the current-density term over the field's 
divergence and can 
apparently lead to different disambiguation 
results, together with a number of other (mentioned but unspecified
by LE2009) keywords in the code. LE2009 used $\lambda =1$.  
 
It becomes, then, a valid question to ask why the ME0 method scores  
better than AZAM and the physics-based methods  
for the limited-resolution flowers cases (Figure 11 
of LE2009). From the methodology of simulated annealing one 
gathers that the method is guaranteed to asymptotically reach the 
global minimum of the functional $\mathcal{F}$  
at an infinite number of iterations  
\cite{metropolis_etal53,press_etal92}. 
Inspecting the functional, its minimum is 
determined by both terms, $|\nabla \cdot \mbf{B}|$ and 
$|J_z|$ ($\lambda =1$). In 
case one term is much larger than the other, however,   
the global minimum of 
$\mathcal{F}$ will be largely dictated by the dominant term. This 
author speculates that $|J_z| \gg |\nabla \cdot \mbf{B}|$ in the 
limited-resolution cases of LE2009, so the disambiguation naturally seeks 
the smoothest possible solution for the horizontal field, that is, the 
one giving rise to the smallest $|J_z|$ over the disambiguation plane, 
largely regardless of $|\nabla \cdot \mbf{B}|$. Physics-based methods, 
on the other hand, enforce $\nabla \cdot \mbf{B} =0$ at each 
iteration. The smoothest possible solution of LE2009 
is merely the one appropriate for this {\it particular}  
flowers case that is potential in full resolution, so ME0 manages to 
reproduce the solution better than the physics-based methods.  
It would be interesting to see how would ME0 score if the test 
structure included electric currents in full resolution and/or was 
placed far from disk center.  
 
Per the above, two questions need to be addressed here:  
\ben 
\item[(1)] Is it appropriate to work with magnetic field data in which 
  $\nabla \cdot \mbf{B} \ne 0$ locally (see Figure 8 of LE2009)?   
  At this point I agree with these authors in that the answer is yes.  
  Solar vector magnetograms include unresolved 
  structure so $\nabla \cdot \mbf{B} = 0$ may be locally violated -  
  we must learn how to handle these data in order to provide 
  disambiguation solutions as close to the divergence-free condition 
  as possible.  
\item[(2)] Are optimization methods - that appear to work even in 
  the finite-volume cases - preferable over physics-based 
  methods? This remains to be determined and is subject to introducing 
  test cases that are more likely to be encountered in the real Sun. This 
  comment targets the narrow validity of the flowers model in 
  LE2009, not the fine-scale structure it includes. This characteristic 
  can hardly be considered realistic for solar magnetic fields and, as such, it 
  disables all physics-based methods by design. Performing better in an unrealistic 
  (finite-size) test case is not proof that optimization methods work better
  than others. Solar-like (semi-infinite) test cases of whatever complexity
  (electric currents, stresses, shear) {\it including} limited spatial
  resolution will unquestionably reveal the best-performing disambiguation
  techniques.  
\een 
 
The computing time required to reach disambiguation results is another 
crucial aspect of this debate: albeit a 
viable concept ever since \inlinecite{metcalf_94} devised it, the ME0 
method, like every simulated annealing technique, is computationally 
intensive and hence inherently slow, significantly slower than other 
techniques. If disambiguation results between different methods are  
so similar that no further benefit exists in additional computing, 
then one will naturally 
opt to use the method reaching these results faster. The same 
will be even more true in case slower optimization methods  
perform slightly worse than other, faster 
methods. In case optimization methods are clearly the best performers 
then, of course, they will be preferable despite their computational 
expense. Which of the three is the case is yet to be seen  
despite the conclusions of LE2009; this work clearly 
shows that further investigation is needed to determine the various 
methods' performance in limited-resolution conditions. 
\subsection{Conclusion} 
\label{S-final} 
This work demonstrates that the disambiguation test for limited 
spatial resolution undertaken by LE2009 was problematic: 
without physical justification it included a synthetic 
magnetic structure with such narrow validity in space (only $0.18''$ 
above the photosphere) that it was unlikely for physics-based 
methods seeking a unique, semi-infinite solution for  
potential and/or non-potential fields to work 
properly. Instead of the actual problem, LE2009 
attributed these methods' failure to 
limited spatial resolution, which was misleading. I showed that 
physics-based disambiguation - even a simplistic, 
potential-field method - applied to a semi-infinite magnetic 
configuration with the same degree of unresolved, fine-scale structure 
reproduces the correct disambiguation solutions almost completely. Moreover,  
I showed that, regardless of fully resolved or 
unresolved structure, physics-based 
methods fail at precisely the areas where the finite-volume approach  
influences transverse fields most heavily,  
forming angles $> \pi/2$ compared to the 
transverse fields of the semi-infinite volume approach.  
 
In conclusion, synthetic test data are often very useful research tools but they 
come at a price: one must ensure that they fulfill to the best possible extent 
the fundamental conditions of the physical system which they are 
designed to reproduce. Otherwise, problems that are not caused by the 
concepts or methodologies employed to analyze the test data, but 
by the test data themselves, are likely to appear. 
 
\section{Closing Remarks}
\label{S-rem}
In a series of 
private communications with K.D. Leka (2009) the author argued  
that the original ``flowers'' case of LE2009 was a 
``perfect storm'' for disambiguation efforts and that if it could be 
used to ``gently nudge'' colleagues away from potential-field 
acute-angle disambiguation, this would be a ``major success''. In 
principle the above are true (I also believe that 
potential-field disambiguation is unrealistic for most 
photospheric conditions, especially those involving multipolar, stressed, or
sheared magnetic structures) but we have a responsibility to show this 
using proper means. A way along these lines  
({\it i.e.}, the semi-infinite ``flowers'' solution) was proposed to the
authors of LE2009 but was not accepted,
which stimulated this work. Based on the results shown 
here, the problem of limited-resolution tests 
and the performance of disambiguation methods in them remains 
to be properly addressed by the community. 

This author is open to further interaction and collaboration aiming to 
address the problems discussed here and to define the state-of-the-art in 
azimuth disambiguation. The semi-infinite ``flowers'' 
solution, in both its ``answer'' and its $180^\circ$-ambiguous versions in 
full and limited spatial resolution is available  
online{\footnote{\url{http://astro.academyofathens.gr/people/georgoulis/data/flowers_semi-infinite_solution/}}} 
for reproduction and validation by the interested researcher. 
%
\begin{acks} 
The author is grateful to N. E. Raouafi and D. M. Rust for numerous 
discussions, encouragement, and valuable feedback on the paper. 
He also thanks K. D. Leka and colleagues for discussing 
his concerns and for making their original synthetic data available. The 
Institute of Space Applications and Remote Sensing (ISARS) of the 
National Observatory of Athens is gratefully acknowledged for the 
availability of their computing cluster facility for runs related to 
this work. During the author's tenure at the Johns Hopkins University 
Applied Physics Laboratory (JHU/APL) in Laurel, MD, USA, this work 
and the author's participation to the Azimuth 
Disambiguation Working Group received partial support from NASA's LWS 
TR\&T grant NNG05GM47G. In its latest stages this work has received support
from the European Union's Seventh Framework Programme (FP7/2007-2013) under
grant agreement n$^\circ$ PIRG07-GA-2010-268245. 
\end{acks} 
%
%
\bibliographystyle{spr-mp-sola} 
\bibliography{SOLA_ms_refs} 
\end{article}  
\end{document}